\documentclass[12pt]{article}
\usepackage{epsfig}
\usepackage{graphicx}
\usepackage{amssymb}
\usepackage{amsmath}
\newcommand{\bra}[1]{{\langle#1|}}

\newcommand\be{\begin{equation}}
\newcommand\ee{\end{equation}}
\newcommand\bea{\begin{eqnarray}}
\newcommand\eea{\end{eqnarray}}

\newcommand\pty{{${\sf PT}$}-symmetry}
\newcommand\ptc{{${\sf PT}$}-symmetric}
\newcommand\pti{{${\sf PT}$}-invariant}
\newcommand\pte{{${\sf PT}$}-invariance}
\begin{document}
\noindent{\large \sf An Algorithmic Test for Diagonalizability
of Finite-Dimensional PT-Invariant Systems}\\ \\
\noindent{\sc \normalsize Stefan Weigert} \\ \\
\noindent{\normalsize \rm $\bra{\mbox{Hu}}\mbox{MP}\rangle$ -
\rm Department of Mathematics, University of Hull\\
UK-Hull HU6 7RX, United Kingdom \\
{\tt S.Weigert@hull.ac.uk}\\
June 2005 
\\ \\
\parbox[height]{13cm}{\small
A non-Hermitean operator does not necessarily have a complete set
of eigenstates, contrary to a Hermitean one. An algorithm is
presented which allows one to decide whether the eigenstates of a
given \pti\ operator on a finite-dimensional space are complete
or not. In other words, the algorithm checks whether a given \ptc\ matrix
is diagonalizable. The procedure neither requires to calculate any
single eigenvalue nor any numerical approximation. \\ \\}
%
%
%

%
\subsection*{\sf I. Introduction}
The physical interpretation of \pti\ operators---and hence their
relevance for the description of physical systems---conti\-nues to
be debated \cite{bender04,mostafazadeh04,weigert04}. There is, however, no doubt about the cathartic role of \pty : it has become more
evident what it means to let go hermiticity in exchange for a
weaker property such as \pte . The success and ease to describe
quantum mechanical systems in terms of hermitean operators is based
on two of their generic properties, namely the existence of 
{\em real} eigenvalues and their {\em diagonalizability}, i.e. the completeness of their orthonormal eigenstates. These properties do not persist if a quantum system was described by a \ptc\ Hamiltonian: its eigenvalues could be complex, and its eigenfunctions would, in general, neither be pairwise orthogonal nor form a complete set. Given a \pti\ operator, it thus appears desirable to decide whether it is diagonalizable or not.

The purpose of this contribution is to provide an algorithm
answering the question of whether a given \pti\ Hamiltonian
operator in a finite-dimensional space {\em does} or {\em does
not} possess a complete set of eigenstates. It is convenient to
represent such an operator as a \ptc\ matrix ${\sf M}$, say. A
procedure will be outlined which, after a finite number of steps,
will announce whether the matrix ${\sf M}$ at hand is 
diagonalizable or not. In principle, the algorithm can be carried
out by hand for matrices of any dimension, and no approximations
are necessary.

Often, the question of diagonalizability will arise in a more general
setting where one considers not just a single matrix but a {\em
family}\ of \ptc\ matrices ${\sf M} (\varepsilon), \varepsilon \in
\mathbb R$. The parameter $\varepsilon$ measures the strength
$\varepsilon \in \mathbb R$ of a ``perturbation'' which destroys hermiticty while repecting \pte . As the parameter varies, all of the cases described previously may occur: typically, two real eigenvalues merge into a single real one at a critical value of $\varepsilon$, subsequently splitting into a pair of two complex conjugate eigenvalues, or {\em vice versa}. These dramatic modifications are accompanied by changes in the nature of the eigenstates of the \pti\ operator, possibly no longer spanning the space on which ${\sf M}(\varepsilon)$ acts.

This behaviour can be understood in terms of so-called {\em
exceptional points} \cite{kato84} which are known to occur 
when a matrix is subjected to the perturbation depending analytically on
a parameter such as $\varepsilon$. At such a point, the
corresponding matrix is not diagonalizable, and its
spectrum may undergo a qualitative change. For a {\em hermitean} operator
subjected to a parameter-dependent {\em hermitean} perturbation, exceptional points cannot occur.

If one applies the algorithm testing for diagonalizability to a
parameter-dependent matrix ${\sf M} (\varepsilon)$, it will output
a polynomial in $\varepsilon$ instead of a number. Its zeros
correspond to those values of the perturbation parameter where the
matrix family ${\sf M} (\varepsilon)$ has exceptional points.
The matrices corresponding to these values of the perturbation
are not diagonalizable, and the spectra of matrices for nearby values of the parameter differ qualitatively.

The following section summarizes the properties of \pti\ systems in
terms of ($2 \times 2$) matrices. Then, the link between
diagonalizability and the so-called minimal polynomial is
reviewed. In Section 3, the algorithmic test is presented
which consists of constructing the minimal polynomial of the matrix followed by a search for degenerate roots by means of the Euclidean algorithm. Various methods are known to effectively
calculate the minimal polynomial of a matrix, outlined in Section
4. Simple examples are studied in Section 5, leading to some
general conclusions about the structure of \ptc\ Hamiltonian
operators in finite-dimensional spaces. Section 6 summarizes the
results and discusses the challenge to extend them to state spaces
of infinite dimension.
 \subsection*{\sf II. \pti\ systems}
A matrix {\sf H} is \pti\ \cite{bender+98}, 
 \begin{equation}\label{ptinv}
 [{\sf H} , {\sf P} {\sf T} ] = 0 \, ,
\end{equation}
if it commutes with the product of parity ${\sf P}$ and the anti-unitary operation of time reversal ${\sf T}$, represented here by complex conjugation, ${\sf T}^\dagger {\sf T} {\sf H} = {\sf H}^*$. 
Eq. (\ref{ptinv}) implies that the characteristic polynomial of any \ptc\ operator {\sf H} has real coefficients only. Consequently, its roots are either real or come in complex-conjugate pairs. One way to show this is to construct a basis in which the Hamiltonian
has real matrix elements only \cite{bender+02}. 

Let us briefly review the properties of \ptc\ systems by
considering the most general \pti\ matrix of dimension $2$,
 \begin{equation}\label{2by2}
 {\sf H} = \left(
          \begin{array}{cc}
           a   & b \\
           b^* & a^*
          \end{array} \right) \, , \quad a,b \in  \mathbb C \, ,
\end{equation}
with parity given by the Pauli matrix $\sigma_x$ in the standard
representation. For real numbers $a$ and $b$, the
matrix ${\sf H}$ is not only \pti\ but also hermitean. Thus, its eigenvalues are {\em real}, and its orthonormal eigenstates span ${\mathbb C}^2$. For $a^* \neq a$ and $b=0$, ${\sf H}$ has a pair of complex conjugate eigenvalues and two orthonormal eigenstates. Matrices of the form
 \begin{equation}\label{interesting}
  {\sf H} = \left(
          \begin{array}{cc}
           i   & b \\
           b & -i
          \end{array} \right) \, ,
          \quad b \in [-1,1] \, ,
\end{equation}
are particularly interesting. For $|b| < 1$, one finds a pair of
two complex conjugate eigenvalues,
 \begin{equation}\label{ccpair}
 E_\pm = \pm \sqrt{b^2-1}  \in i \mathbb R \, ,
\end{equation}
associated with two non-orthogonal eigenstates,
 \begin{equation}\label{nononeigenstates}
 \frac{1}{\sqrt{2} b }\left(
 \begin{array}{c}
 b \\
 i - \sqrt{b^2-1}
 \end{array} \right)
 \, , \qquad
  \frac{1}{\sqrt{2} b }\left(
 \begin{array}{c}
 b \\
 i + \sqrt{b^2-1}
 \end{array} \right)
 \, .
\end{equation}
When $b = \pm 1 $ in (\ref{interesting}), ${\sf H}$ has a two-fold
degenerate eigenvalue, $E_0=0$, and there is only {\em one}
eigenstate, namely,
 \begin{equation}\label{singleeigenstate}
 \frac{1}{\sqrt{2}}\left(
 \begin{array}{c}
 \mp 1 \\
 i
 \end{array} \right)
 \, .
\end{equation}
This situation, impossible for a hermitean matrix, is usually
described by saying that the {\em algebraic} multiplicity of the
eigenvalue $E_0$ is two while its {\em geometric} multiplicity
equals one: the characteristic polynomial of $\sf M$ has a double root 
associated with a single eigenvector only. In this case, the matrix ${\sf H}$ is not diagonalizable: a similarity transformation sending it 
to a {\em diagonal} matrix cannot exist since its eigenstates {\em would} span the space ${\mathbb C}^2$.
\subsection*{\sf III. Diagonalizability and the minimal polynomial of a matrix}
Each square matrix $\sf M$ of dimension $N$ satisfies the identity
\begin{equation}
p_{\sf M}( {\sf M}) = 0 \, ,
\label{CayleyHamM}
\end{equation}
where $p_{\sf M}(\lambda)$ is the {\em characteristic polynomial} of ${\sf M}$,
\begin{equation}\label{charpolM}
p_{\sf M}( \lambda )
         = \det \left( \lambda {\sf E} - {\sf M} \right) \, ,
\ee
with ${\sf E}$ being the unit matrix of dimension $N$. In other
words, the characteristic polynomial of ${\sf M}$ {\em
annihilates} the matrix ${\sf M}$. The polynomial $p_{\sf
M}(\lambda)$ has degree $N$ and it is a {\em monic} polynomial,
that is, the coefficient multiplying the highest power of
$\lambda$ is equal to $1$. Obviously, many other monic
polynomials of {\em higher} degree also annihilate ${\sf
M}$: simply take $p^2_{\sf M}(\lambda), p^3_{\sf M}(\lambda),
\dots$ It is less obvious, however, whether one can find
polynomials of degree {\em less} than $N$ which annihilate ${\sf
M}$. This, in fact, depends on the properties of the matrix ${\sf
M}$.

Define \cite{lancaster+85} the {\em minimal
polynomial} of the matrix {\sf M} as the monic polynomial
$m_{\sf M}(\lambda)$ of {\em least} degree which
annihilates {\sf M}:
\begin{equation}\label{minpolM}
m_{\sf M}( \sf M ) = 0 \, .
\ee
The minimal polynomial $m_{\mbox{\small $\sf M$}}( \lambda)$ is unique \cite{lancaster+85}, and its degree $N_0$ is less than or equal to the degree of the characteristic polynomial, $N_0 \leq N$. The minimal polynomial divides the characteristic polynomial without a remainder, $m_{\sf M}( \lambda) \left. \right|  p_{\sf M}(\lambda)$, or equivalently,
\begin{equation}\label{dividecharpol}
p_{\sf M}( \lambda)
   = d_{\sf M}(\lambda) m_{\sf M}( \lambda) \, ,
\ee
where $d_{\sf M}(\lambda)$ is a non-zero polynomial of degree less than
$N$. The characteristic and the minimal polynomial of the matrix
{\sf M} coincide in the case where $d_{\sf M}(\lambda) \equiv 1$.

In general, the minimal polynomial has $\nu_0$ roots $M_\nu$,
\begin{equation}\label{genminpoly}
m_{\sf M} (\lambda)
 = \prod_{\nu=1}^{\nu_0} (\lambda - M_\nu)^{\mu_\nu} \, ,
   \quad \nu_0 \leq N \, ,
\ee
with multiplicities $\mu_\nu$ summing to $N_0 = \mu_1 + \mu_2 + \dots + \mu_{\nu_0}$. Here is the important property of the polynomial $m_{\sf M} (\lambda)$:  the matrix  {\sf M} is diagonalizable if and only if each root $M_\nu$ in (\ref{genminpoly}) has multiplicity one, $\mu_\nu
\equiv 1, \nu =1, \dots, \nu_0$, that is,
\begin{equation}\label{genminpolydiag}
m_{\sf M}( \lambda)
 = \prod_{\nu=1}^{\nu_0} (\lambda - M_\nu)\, ,
   \quad \mbox{all } \, M_\nu \mbox{ distinct} \, .
\ee
No polynomial of degree less than $m_{\sf M}( \lambda)$  annihilates the matrix ${\sf M}$.

Let us illustrate the properties of minimal polynomials using
low-dimensional matrices. Consider the matrix {\sf A}
with entries $(1,1,2)$ on the diagonal, and zero elsewhere. Its characteristic polynomial is given by
\begin{equation}\label{charpolA}
p_{\sf A}(\lambda)= (\lambda-1)^2 (\lambda-2) \, ,
%
\ee
while its minimal polynomial reads
\begin{equation}\label{minpolA}
m_{\sf A}(\lambda)= (\lambda-1) (\lambda-2) \, ,
%
\ee
being of the form (\ref{genminpoly}), with $N_0 = \nu_0 = 2$.
This is easy to verify since $m_{\sf A}(\sf A) = {\sf A}^2 -3 {\sf A}+2{\sf E}=0 $ holds, while none of its factors annihilates ${\sf M}$: both $({\sf M} - {\sf E}) $ and $({\sf M} - 2{\sf E})$ are different from zero. Thus, the minimal polynomial divides the characteristic one, $p_{\sf A}(\lambda) = (\lambda - 1) m_{\sf A}(\lambda)$, leading to $d_{\sf A}(\lambda) = (\lambda -1)$. Due to (\ref{minpolA}), the matrix ${\sf A}$ {\em is} diagonalizable--- a correct but hardly surprising result since the matrix {\sf A} has been diagonal from the outset.

Here is the instructive part of the example: consider the matrix
\begin{equation}\label{defB}
{\sf B} = \left(
          \begin{array}{ccc}
                1 & b & 0 \\
                0 & 1 & 0 \\
                0 & 0 & 2
          \end{array}
          \right) \, , \qquad b \in \mathbb C  \, ,
\ee
which is different from {\sf A} as long as $b$ is different from
zero. The characteristic polynomial of {\sf B} equals that of {\sf A} but the matrix {\sf B} must have a different
minimal polynomial since $m_{\sf A}({\sf B}) \neq 0$. It is not difficult to verify that no linear or quadratic polynomial annihilates {\sf B} as long as $b \neq 0$. This implies that its minimal polynomial coincides with its characteristic polynomial,
\begin{equation}
p_{\sf B}(\lambda)
 = m_{\sf B}(\lambda) \, , \quad d_{\sf B}(\lambda) \equiv 1 \, .
\label{Bpolys}
\ee
Consequently, the minimal polynomial of {\sf B} does {\em not} have the form specified in (\ref{genminpolydiag}), and the matrix ${\sf B}$ is {\em not} similar to a diagonal matrix. Inspection shows that {\sf B} indeed contains a $(2 \times2)$ Jordan block for any nonzero value of $b$. 

For \pti\ matrices, both the polynomials $d_{\mathsf{H}}(\lambda)$ and $m_{\mathsf{H}}(\lambda)$ have real coefficients only, just as the characteristic polynomial. This will be shown once the function $d_{\mathsf{M}}(\lambda)$ in (\ref{dividecharpol}) has been defined in general (cf. Sec. {\sf IV.2}).
\subsection*{\sf IV. An algorithmic test for diagonalizability}
A square matrix ${\sf M}$ of dimension $N$ is diagonalizable  if its minimal polynomial is a product of factors $(\lambda - M_\nu)$ with all numbers $M_\nu, \nu = 1,2, \ldots , n \leq \nu_0$, distinct, as shown in Eq. (\ref{genminpolydiag}). Consequently, to test for diagonalizability of given a matrix {\sf M}, one needs to
\begin{enumerate}
\item[($\imath$)] find the minimal polynomial $m_{\sf M}(\lambda)$ of the matrix {\sf M};
\item[($\imath \imath$)] determine whether the polynomial $m_{\sf M}(\lambda)$ has single roots only.
\end{enumerate}
To calculate numerically the roots of either the characteristic or the minimal polynomial is not a valid approach since, in general, the {\em exact} roots of a polynomial cannot be specified in a finite procedure. Any algorithmic implementation must generate answers to ($\imath$) and ($\imath \imath$) in a {\em finite} number of steps. Note that even if the first step has been implemented, it is unlikely that the minimal polynomial will emerge in factorized form. 

Interestingly, it is possible to construct the minimal polynomial of a matrix and to check for degenerate roots in a finite number of steps. In both cases one searches for common factors of polynomials, which is achieved algorithmically by the {\em Euclidean division algorithm}. These results seem to have been put together for the first time in \cite{abate97} in order to decide algorithmically whether a given matrix is diagonalizable. As it stands, it could be applied to the non-hermitean matrix govering the motion of two coupled damped classical oscillators studied in \cite{heiss04}. 

In the following, a slightly simplified approach to the problem of diagonalizability is presented, adapted to matrices with \pty. Before
implementing the steps ($\imath$)  and ($\imath\imath$), the Euclidean algorithm for polynomials will be presented briefly to establish notation.
\subsubsection*{\sf IV.1 Euclidean division algorithm for polynomials}
Given two integer numbers $p_0 > p_1$, say, the Euclidean division
algorithm outputs their greatest common divisor, denoted by
$\mbox{gcd}(p_0,p_1) \in {\mathbb N}_0$, after a finite number of
steps. It works as follows: first, you need to express the larger
number as $q_1$-fold multiple of the smaller number plus a
remainder $p_2$,
\begin{equation}\label{eucl}
p_0 = q_1 p_1 + p_2 \, , \qquad q_1,p_2 \in {\mathbb N}_0\, ,
                 \quad   p_1 > p_2 \leq 0 \, .
\ee
This relation implies that any common divisor of $p_0$ and $p_1$ divides $p_2$ as well, hence $\mbox{gcd}(p_0,p_1) = \mbox{gcd}(p_1,p_2)$. Thus, it is sufficient to search for the greatest common divisor of the pair $(p_1,p_2)$. This can be achieved by increasing each index in (\ref{eucl}) and feeding in the pair $(p_1,p_2)$ instead of $(p_0,p_1)$, etc. Since $p_0>p_1$ and $p_1>p_2$, the algorithm will stop after a finite number of iterations and produce a remainder equal to zero, $p_{k+1}=0$, say.  The {\em non-zero} remainder $p_k$ generated in the penultimate step is the desired result, $\mbox{gcd}(p_0,p_1) = p_k$. If $\mbox{gcd}(p_0,p_1)= 1$, the numbers $p_0$ and $p_1$ are relatively prime, otherwise a common divisor different from one has been identified.

A polynomial in the variable $\lambda$ can be written as a unique
product of linear factors $(\lambda - \lambda_n)$ where the
numbers $\lambda_n \in \mathbb C$ are its roots. This
representation makes polynomials similar to integer numbers in
some respects. The equivalent of the Euclidean algorithm, when
applied to two polynomials, outputs their greatest common divisor,
which is a polynomial itself. This result is based on the fact
that any two polynomials $p_0(\lambda)$ and $p_1(\lambda)$, with
$\deg p_0(\lambda) > \deg p_1(\lambda)$ are related by
\begin{equation}\label{euclpoly}
p_0(\lambda) = q_1(\lambda) p_1(\lambda) + p_2(\lambda) \, ,
              \quad \deg p_1(\lambda) > \deg p_2(\lambda) \geq 0 \, ,
\ee
which is the equivalent of (\ref{eucl}). The polynomials $q_1(\lambda)$, with $\deg q_1(\lambda) = (\deg p_0(\lambda) - \deg p_1(\lambda))$, and hence $p_2(\lambda)$, are  found from long division. If $p_0(\lambda)$ and $p_1(\lambda)$ have
a common factor, then $p_2(\lambda)$ must have this factor as well. Thus, it is sufficient to search for $\mbox{gcd}(p_1(\lambda),p_2(\lambda))$ instead of $\mbox{gcd} (p_0(\lambda),p_1(\lambda))$ but the degrees of the polynomials involved have effectively been reduced. Consequently, this procedure can be repeated all over again and it halts once a {\em vanishing} remainder has been obtained, $p_{k+1}(\lambda) = 0$, say. Then, the greatest common factor of the polynomials $p_0(\lambda)$ and $p_1(\lambda)$ is
given by the last non-zero remainder polynomial, $p_k (\lambda)$, calculated  in the next-to-last application of the algorithm. If $\deg p_k (\lambda) = 0$, the initial polynomials are ``relatively prime,'' otherwise their greatest common divisor is a polynomial of degree at least one.
\subsubsection*{\sf IV.2 Step ($\imath$): Finding the minimal polynomial of a matrix}
The function $d_{\sf M} (\lambda)$ relates the minimal polynomial $m_{\sf M}(\lambda)$ of the matrix {\sf M} to its characteristic polynomial $p_{\sf M}(\lambda)$ according to Eq.
(\ref{dividecharpol}). Hence, the minimal polynomial associated with ${\sf M}$ is known once the characteristic polynomial and the function $d_{\sf M} (\lambda)$ have been determined.

Two steps are required to construct the function $d_{\sf M} (\lambda)$ \cite{lancaster+85}. First, you need to calculate the matrix ${\sf D}_{\sf M} = \mbox{adj } (\lambda {\sf E} - {\sf M})$, given by the transposed cofactors---or signed minors---of the matrix $(\lambda {\sf E} - {\sf M})$. The adjoint of a matrix, {\sf C} say, always exists, and it satisfies the relation
 \begin{equation}\label{adjoint}
 {\sf C} \,  \mbox{adj } {\sf C} = (\det {\sf C}) {\sf E} \, .
\end{equation}
For $\det \sf C \neq 0$, Eq. (\ref{adjoint}) leads to the familiar expression of the inverse matrix of ${\sf C}$.

According to \cite{lancaster+85} the polynomial $d_{\sf M}(\lambda)$ is given by the  {\em greatest} (monic) {\em common divisor} of the $N^2$ elements of $\mbox{adj } (\lambda {\sf E} - {\sf M})$,
 \begin{equation}
 d_{\mathsf{M}} (\lambda) 
  = \mbox{gcd} \left\{ (\mathsf{D}_{\sf M})_{nm} |                       
                                       n,m=1,\ldots,N\right\}   
   \, , 
 \label{dhldefinition}
 \end{equation}
Thus, in a second step, you need to apply the Euclidean algorithm to all pairs of entries of the matrix $\mbox{adj } (\lambda {\sf E} - {\sf M})$. Having thus identified the function $d_{\sf M} (\lambda)$, the minimum polynomial of {\sf M} follows from (\ref{dividecharpol}),
 \begin{equation}\label{findminpoly}
 m_{\sf M} (\lambda)
   = \frac{p_{\sf M} (\lambda)}{d_{\sf M}(\lambda)} \, .
\end{equation}

Now it is possible to show that, for a \pti\ matrix, the polynomials $d_{\sf M}(\lambda)$ and $m_{\sf M} (\lambda)$ have real coefficients only, just as the characteristic polynomial. Using a basis in which all elements of {\sf H} are real, leads to 
 \begin{equation}
 ({\sf D}_{\sf H} (\lambda))^* 
  = \left( \mbox{adj}(\lambda \mathsf{E} - \mathsf{H} ) \right)^{*} 
  = \mbox{adj} (\lambda^{*} \mathsf{E}-\mathsf{H)} 
  = {\sf D}_{\sf H} (\lambda^*) \, . 
 \label{realadjoint}
 \end{equation}
 which states that in this basis the adjoint of {\sf H} has only real matrix elements (except for the unknown $\lambda$). Taking (\ref{dhldefinition}) into account this leads to 
 \begin{equation}
 \left( d_{\mathsf{H}} (\lambda) \right)^{*}
 =\mbox{gcd} \left\{ (\mathsf{D}_{\mathsf{H}})_{nm}(\lambda^{*})|n,m=1,\ldots,N\right\}
 = d_{\mathsf{H}} (\lambda^{*}) \, ,
 \label{dhlreaL}
 \end{equation}
which, in conjunction with $\left( p_{\mathsf{H}} (\lambda) \right)^{*}    = p_{\mathsf{H}} (\lambda^{*})$ and Eq. (\ref{findminpoly}) implies indeed $\left( m_{\mathsf{H}}(\lambda) \right)^{*}=m_{\mathsf{H}}(\lambda^{*})$.

Let us verify that this procedure outputs the correct minimal polynomials for the matrices {\sf A} and {\sf B} introduced in Section 3. The adjoint of the matrix $(\lambda {\sf E} - {\sf B})$ reads
 \bea \label{cofB}
 \mbox{adj } (\lambda {\sf E} - {\sf B})
       &=& \mbox{adj } \left(
         \begin{array}{ccc}
            \lambda - 1 & - b & 0 \\
            0 & \lambda -1 & 0 \\
            0 & 0 & \lambda -2
          \end{array}
         \right) \nonumber \\
        &=& \left(
         \begin{array}{ccc}
            (\lambda - 1) (\lambda -2) & b(\lambda -2) & 0 \\
            0 &  (\lambda - 1) (\lambda -2) & 0 \\
            0 & 0 &   (\lambda - 1)^2
          \end{array}
         \right) \, .
\eea
Due to the simplicity of the matrices involved, the Euclidean algorithm can be run ``by inspection:'' for $b\neq 0$, the only common factor  among the entries in (\ref{cofB}) is given by $d_{\sf B}(\lambda) = 1$. Consequently, the minimal and the characteristic polynomial of {\sf B} coincide as stated in Eq. (\ref{Bpolys}). If the parameter $b$ takes the value zero, {\sf B} turns into {\sf A}, and and a non-constant greatest common divisor emerges, $d_{\sf A}(\lambda) = (\lambda - 1)$. Using Eq. (\ref{dividecharpol}), one obtains the minimal polynomial $m_{\sf A}(\lambda) = (\lambda -1)(\lambda-2)$, agreeing with Eq. (\ref{minpolA}).

In \cite{abate97}, a different approach to determine the minimal
polynomial of a matrix {\sf M} has been presented which, ultimately, is also based on finding the greatest common divisor of specific  polynomials. According to \cite{horn+99}, any method to determine whether the matrices ${\sf M}^0 \equiv {\sf E}, {\sf M}, {\sf M}^2, \ldots , {\sf M}^{N-1}$, are linearly dependent, can be used to construct the minimal polynomial of {\sf M}; two such methods are described in this reference, and a third one can be found in \cite{horn+85}. The latter approaches have in common that they are {\em not} based on the Euclidean algorithm. For actual calculations, it is convenient to resort to a {\tt Mathematica} program  \cite{weisstein+05} to find the minimal polynomial of a matrix ${\sf M}$.
\subsubsection*{\sf 4.3 Step ($\imath \imath$): Identifying degenerate
roots of a polynomial}
Once the minimal polynomial $m_{\sf M} (\lambda)$ has been found,
one needs an algorithm to decide whether it has single roots only
\cite{abate97}. Imagine a polynomial $m(\lambda)$ to have an
$s$-fold root $\lambda_0$, $2 \leq s \leq N$. Its factorization
reads
 \begin{equation}\label{sfold}
 m(\lambda) = (\lambda - \lambda_0)^s \ldots \, ,
\end{equation}
where the dots indicate a polynomial of degree $(N-s)$. Its
derivative takes the form
 \begin{equation}\label{sfoldprime}
 \frac{dm}{d\lambda} = (\lambda - \lambda_0)^{s-1} \dots  \, ,
\end{equation}
 the dots standing again for some polynomial of degree $(N-s)$.
Obviously, the polynomial and its  derivative are not relatively
prime: $m(\lambda)$ and $m^\prime(\lambda)$ have a factor
$(\lambda - \lambda_0)^{s-1}$ of order at least one in common.
Thus, applying the division algorithm to the pair $(m_{\sf
M}(\lambda), m_{\sf M}^{\prime} (\lambda))$ checks whether the
polynomial $m_{\sf M} (\lambda)$ has the form
(\ref{genminpolydiag}). If the procedure outputs $\mbox{gcd}
(m_{\sf M}(\lambda), m_{\sf M}^{\prime} (\lambda)) \propto 1$, all
roots of $m_{\sf M}$ are distinct and the associated matrix {\sf
M} is diagonalizable, otherwise it is not.

This concludes the description of an algorithm to test for
diagonalizability of a given \ptc\ matrix ${\sf M}$. No
fundamental changes are necessary if one studies a
parameter-dependent family of matrices ${\sf M} (\varepsilon)$. However, the algorithm will output conditions polynomial
in the parameter $\varepsilon$, indicating specific parameter
values where diagonalizability breaks down. It is convenient to
study the resulting modifications by working out some simple
examples, illustrating at the same time the proposed algorithm.

 \subsection*{\sf V. Examples}
 \subsubsection*{\sf V.1 Matrices of dimension ($2\times 2$)}

Let us apply the algorithm described above to the matrix ${\sf H}$
in (\ref{2by2}}) assuming the numbers $a$ and $b$ to be different
from zero. Its characteristic polynomial reads
 \begin{equation}\label{charpolH}
 p_{\sf H} (\lambda)
    = \det (\lambda {\sf E} - {\sf H})
    = \lambda^2 - 2 (\Re a) \lambda + | a |^2  - |b|^2 \, ,
\end{equation}
while its minimal polynomial is found via the function $d_{\sf
H}(\lambda)$ equal to the highest common factor of the matrix
 \begin{equation} \label{hcfH?}
 {\sf D}_{\sf H} (\lambda) =  \mbox{adj } (\lambda {\sf E} - {\sf H} )
        = \left(
          \begin{array}{cc}
          \lambda - a^* &  -b \\
          -b^* & \lambda- a
          \end{array}
          \right) \, .
\end{equation}
By inspection, a non-constant factor only exists among the four entries of  ${\sf D}_{\sf H} (\lambda)$ if $b=\Im a =0$. In this case, ${\sf H}$ turns into a real multiple of the identity, hence it is diagonalizable. This observation illustrates a fine point of the construction of the minimal polynomial: even upon identifying a non-constant function $d_{\sf H} (\lambda)$, the minimal polynomial $m_{\sf H}(\lambda)$ may still be of the form (\ref{genminpolydiag}). For $b=\Im a =0$, the characteristic polynomial turns into $(\lambda - \Re a)^2$, implying indeed $m_{\sf H} (\lambda) = (\lambda - \Re a)$. Here, the function $d_{\sf H}(\lambda) = (\lambda - \Re a)$ removes factors of the characteristic polynomial which stem from the {\em degeneracy} of an eigenvalue of ${\sf H}$.

From now on, either $b$ or $\Im a$ are assumed to be different from zero, hence $d_{\sf H} (\lambda) =1$, and the minimal polynomial $m_{\sf H}(\lambda)$ is given by Eq. (\ref{charpolH}),
 \begin{equation}\label{char=miniH}
 m_{\sf H} (\lambda) =  p_{\sf H} (\lambda) \, ,
\end{equation}
which concludes the first step of the algorithm.

In the second step of the algorithm, the search for multiple roots of $m_{\sf H}(\lambda) \equiv p_0(\lambda)$, one needs to determine the highest common factor of the minimal polynomial and its derivative,
 \begin{equation}\label{derminpoly}
 m^\prime_{\sf H} (\lambda)
    = 2 (\lambda -  \Re a) \equiv p_1 (\lambda) \, .
\end{equation}
Applying the Euclidean algorithm to the pair $(m_{\sf H}
(\lambda), m^\prime_{\sf H} (\lambda))$ means to solve for a
polynomial $q_1 (\lambda)= A \lambda + B$ and for $p_2 (\lambda)$
such that
 \begin{equation}\label{mmprimeeuclid}
 m_{\sf H} (\lambda)
      = (A \lambda +B) m_{\sf H}^\prime (\lambda) + p_2(\lambda)   \, ,
      \quad A, B \in \mathbb R \, .
\end{equation}
The unknowns are easily obtained as
 \begin{equation}\label{euclappl1}
 A = \frac{1}{2} \, , \quad B= - \frac{1}{2} \Re a \, ,
 \quad p_2(\lambda) = (\Im a)^2 - | b |^2  \simeq \lambda^0 \, .
\end{equation}
Two possibilities now arise: either $p_2(\lambda)$ equals zero or
it does not. The first case occurs if
 \begin{equation}\label{condp2}
 (\Im a)^2 = | b |^2 \, ,
\end{equation}
and the algorithm comes to a halt. As mentioned above, the
greatest common divisor of the initial polynomials is then given
by the penultimate (monic) remainder polynomial, i.e. $p_1
(\lambda) = ( \lambda - \Re a)$. It follows that $m_{\sf H}
(\lambda)$ and its derivative do have a common non-constant
divisor, so that ${\sf H}$ cannot be brought to diagonal form. It
is easy to verify that $m_{\sf H} (\lambda) = ( \lambda - \Re
a)^2$ when (\ref{condp2}) holds, confirming that the minimal
polynomial of ${\sf H}$ has a {\em double} root $\Re a$.
Furthermore, a simple calculation shows that the matrix $\sf H$
has indeed a {\em single} eigenstate only if the relation $\Im a =
\pm |b|$ holds. Note that this result covers the example of a
non-diagonalizable ${\sf H}$ of Section 2, where $a=i$ and $b=\pm
1$ had been considered.

Finally, if $a$ and $b$ do not satisfy (\ref{condp2}), the
remainder polynomial $p_2(\lambda)$ does not vanish. Being a constant, the algorithm is bound to stop after the next iteration. Determine $q_2(\lambda)=(C\lambda + D)$ and $p_3
(\lambda)$ such that
 \begin{equation}\label{mmprimeeuclid2}
  (\lambda -  \Re a)
      = (C \lambda + D) ((\Im a)^2 - | b |^2) + p_3 (\lambda) \, ,
      \quad C, D \in \mathbb R \, ,
\end{equation}
holds, i.e.,
 \begin{equation}\label{euclappl2}
       C = \frac{1}{(\Im a)^2 - | b |^2} \, ,
 \quad D = \frac{- \Re a}{(\Im a)^2 - | b |^2} \, ,
 \quad p_3(\lambda) = 0 \, .
\end{equation}
The algorithm halts indeed due to $p_3(\lambda) = 0$, and the
penultimate remainder is $p_2 (\lambda) \propto 1$, indicating
that the minimal polynomial does not have any degenerate roots,
and ${\sf H}$ is diagonalizable.

In summary, the matrix {\sf H} is diagonalizable for all parameter
values except when $\Im a = \pm |b|$. In this case the algebraic
multiplicity of its eigenvalue is two, while its geometric
multiplicity is one; otherwise the multiplicities both equal two.
It is important to note that it was not necessary at any stage to
determine the eigenvalues of ${\sf H}$.
 \subsubsection*{\sf V.2 Matrices of dimension ($4 \times 4$)}
It is instructive to apply the algorithm to a \pti\ matrix of dimension $4$,
 \begin{equation}\label{4by4}
 {\sf H} = \left(
          \begin{array}{cccc}
          i \varepsilon & s & 0 & 0\\
          s & -i \varepsilon & \delta & 0 \\
          0 & \delta & i \varepsilon & s  \\
          0 & 0 & s & -i \varepsilon \\
          \end{array} \right) \, , \quad
                       s,\delta > 0 \, ,
\end{equation}
which depends on a perturbation parameter $\varepsilon$. As before, the action of ${\sf T}$ on a matrix effects complex conjugation of its entries, while ${\sf P}$ is now given by a $(4 \times 4)$ matrix with entries equal to one along its minor
diagonal and zero elsewhere. Eq. (\ref{ptinv}) is then readily
verified. The characteristic polynomial of ${\sf H}$ is given by
 \begin{equation}\label{charpoldim4}
 p_{\sf H} (\lambda) = \lambda^4 + \alpha_\varepsilon \lambda^2 + \beta_\varepsilon \, ,
\end{equation}
with $\alpha_\varepsilon = 2 \varepsilon^2 - 2 s^2 - \delta^2$ and
$ \beta_\varepsilon = \varepsilon^4 - (2s^2  +
\delta^2)\varepsilon^2 + s^4$. The minimal polynomial of ${\sf H}$
coincides with the characteristic one, $m_{\sf H} (\lambda) \equiv
p_{\sf H} (\lambda)$, since the only common factor of the matrix elements
of ${\sf D}_{\sf H}$ is equal to one, $d_{\sf H} (\lambda) = 1$. To see this, it is sufficient to calculate the two matrix elements $[{\sf D}_{\sf H}]_{14} = -s^2 \delta$, and $[{\sf D}_{\sf H}]_{23} = - (\lambda^2 + \varepsilon^2) \delta$, for
example. Whatever the value of $\varepsilon$, for nonzero $s$ and
$\delta$ the only common divisor is one, so that $d_{\sf H}
(\lambda) =1$.

Let us now determine $\mbox{gcd} (m_{\sf H} (\lambda),
m^\prime_{\sf H} (\lambda ))$ by the Euclidean algorithm, where
$m_{\sf H} (\lambda) \equiv p_0(\lambda) $ is given in Eq.
(\ref{charpoldim4}) and $m^\prime_{\sf H} (\lambda ) \equiv
p_1(\lambda) = 4 \lambda^3 + 2\alpha_\varepsilon\lambda$. Comparing powers of
$\lambda$ in Eq. (\ref{mmprimeeuclid}) with the polynomials just
defined, one obtains
 \begin{equation}\label{edastep2}
 A= \frac{1}{4}\, , \quad B= 0\, ,
 \quad p_2(\lambda) = \frac{\alpha_\varepsilon}{2}\lambda^2 + \beta_\varepsilon \, .
\end{equation}
The algorithm only stops here if $\alpha_\varepsilon =
\beta_\varepsilon = 0$ which would require $s=\delta = 0$, contrary to both $s$ and $\delta$ being different from zero. If $\alpha_\varepsilon=0$ is assumed, ${\sf H}$ is diagonalizable for all $\varepsilon$ since $\beta_\varepsilon$ cannot take the value zero, and the algorithm stops after the next step, outputting $\beta_\varepsilon \propto 1$ as greatest common factor. Assume now $\alpha_\varepsilon \neq 0$ and apply the division algorithm to the pair $(p_1(\lambda),p_2
(\lambda))$. The unknown constants in $q_2(\lambda) = C\lambda +
D$, and the remainder polynomial $p_3(\lambda)$ are found to be
 \begin{equation}\label{edastep3}
 C = \frac{8}{\alpha_\varepsilon}, \quad D = 0 \, ,
     \quad p_3 (\lambda) = \frac{2}{\alpha_\varepsilon}
             (\alpha_\varepsilon^2 - 4 \beta_\varepsilon)\lambda \, .
\end{equation}
For the algorithm to stop, one must have $p_3(\lambda) = 0$. This, however, does not happen whatever the value of $\varepsilon$ since $\alpha_\varepsilon^2 - 4 \beta_\varepsilon = \delta^2 (4s^2 + \delta^2) >0$. The next iteration of the algorithm leads to
 \begin{equation}\label{edastep4}
 E = \frac{\alpha^2_\varepsilon}{4(\alpha_\varepsilon^2 - 4 \beta_\varepsilon)}, \quad F = 0 \, ,
     \quad p_4 (\lambda) = \beta_\varepsilon \, ,
\end{equation}
where $q_3(\lambda) = E\lambda + F$. Producing a remainder polynomial of degree zero in $\lambda$, the condition for the minimal polynomial to have multiple roots is finally given by
 \begin{equation}\label{remaindercond}
 \beta_\varepsilon =  \varepsilon^4 - (2s^2  + \delta^2)\varepsilon^2 + s^4 = 0 \, .
\end{equation}
This fourth-order polynomial in $\varepsilon$ has roots
 \begin{equation}\label{roots}
 \pm \varepsilon_\pm = \pm \sqrt{ \sigma \pm \Delta \sigma} \, ,
  \quad \sigma = s^2 +\frac{\delta^2}{2} > 0 \, ,
  \quad \Delta \sigma = \sqrt{\sigma^2 -s^4} \in (0,\sigma) \, .
\end{equation}
For each of these four real values of the parameter $\varepsilon$,
the matrix ${\sf H}$ is {\em not} diagonalizable. In other words,
the algebraic and geometric multiplicity of the eigenvalues of
${\sf H}$ do not coincide, and its eigenstates do not span the
space ${\mathbb C}^4$. It is important to note that the
eigenvalues of the matrix $\sf H$ are not known at this stage.
\subsubsection*{\sf V.3 The global structure of ${\sf H}$}
\label{sec:53SfGlobalStructure}
Let us now determine the global properties of ${\sf H}$. This is easily done upon combining (\ref{roots}) with the characteristic polynomial (\ref{charpoldim4}) in its factorized form,
 \begin{equation}\label{charpoldim4factorized}
 p_{\sf H} (\lambda) =
 (\lambda - \lambda_{+})(\lambda + \lambda_{+})
 (\lambda - \lambda_{-})(\lambda + \lambda_{-}) \,
\end{equation}
with roots
 \begin{equation}\label{fourevs}
 \pm \lambda_{\pm}
   = \pm \sqrt{\sigma \pm \Delta \sigma - \varepsilon^2} \, ,
\end{equation}
expressed directly in terms of $\sigma$ and $\Delta \sigma$.

To graphically represent the parameter space of {\sf H} and its properties, it is convenient to
eliminate the parameter $s$ by the scaling $\varepsilon \to s
\varepsilon$ and $\delta \to s \delta$. This effectively amounts
to sending $s \to 1$, and Eq. (\ref{remaindercond}) simplifies to
$\varepsilon^4 - 2(1 + \delta^2/2)\varepsilon^2 + 1 = 0$, plotted in 
Fig. 1. 

\begin{figure}[t]
 \begin{center}
	\includegraphics{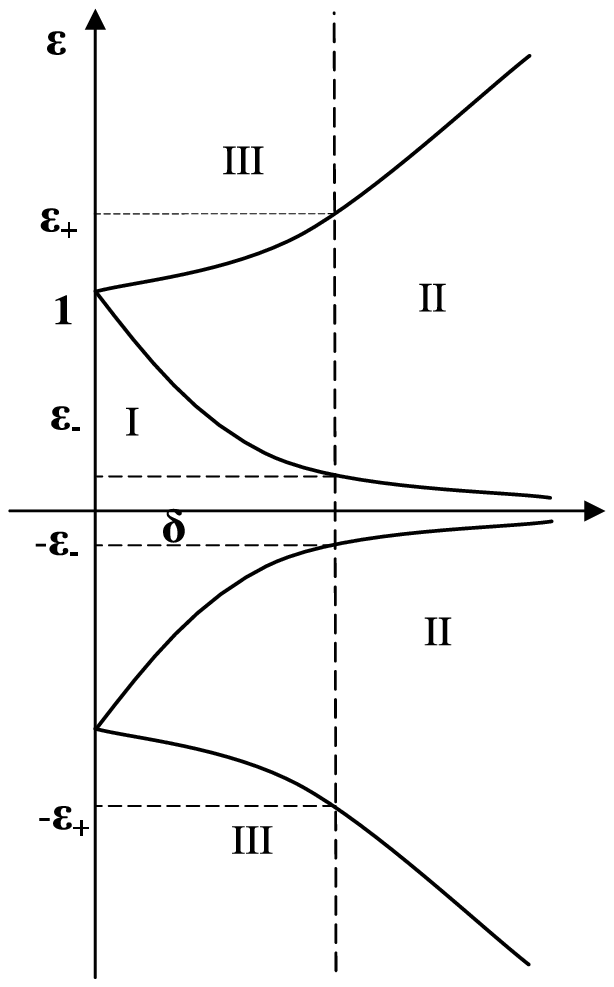}
 \end{center}
	\caption{Parameter Space of the matrix {\sf H} defined in (\ref{4by4}); region I: four real eigenvalues; regions II: two real and one pair of complex-conjugate eigenvalues; region III: a two pairs of complex-conjugate eigenvalues; the matrix {\sf H} is not diagonalizable on the full lines separating the regions I,II, and III}
	\label{parameterspace}
\end{figure}

Imagine to move along the dashed vertical line, determined by a fixed positive value of $\delta>0$ and variable $\varepsilon$. For $\varepsilon = 0$, the matrix ${\sf H}$ is hermitean, hence it has four distinct real eigenvalues and four orthonormal eigenstates. In region I, where $0 < |\varepsilon| < \sqrt{\sigma - \Delta \sigma}$, Eq. (\ref{fourevs}) says that the eigenvalues remain real and distinct; a complete, not necessarily orthonormal set of four eigenstates continues to exist since (\ref{remaindercond}) does not hold. When $\varepsilon = \pm\sqrt{(\sigma - \Delta \sigma)}$, two eigenvalues coincide numerically, and the corresponding two eigenstates merge into a single one, leaving ${\sf H}$ with an incomplete basis. Then, for $\sqrt{\sigma - \Delta \sigma} < |\varepsilon|
< \sqrt{\sigma + \Delta \sigma}$, in region II, two real eigenvalues and a pair of complex-conjugate ones exist, with $\sf H$ being diagonalizable
throughout since (\ref{remaindercond}) does not hold. At
$\varepsilon = \pm\sqrt{(\sigma + \Delta  \sigma)}$, the remaining two
real eigenvalues degenerate to a single one, leaving ${\sf H}$
non-diagonalizable again with only three eigenstates. Finally, in region III, defined by $\sqrt{\sigma + \Delta \sigma} <
|\varepsilon|$, the matrix $\sf H$ is diagonalizable and it comes with two  pairs of complex-conjugate eigenvalues.

Finally, for $\delta=0$, the matrix ${\sf H}$ in (\ref{4by4}) decouples into an pair of identical two-dimensional matrices. The left boundary of region I sees the real eigenvalues of ${\sf H}$ degenerate pairwise which is consistent with the observations made earlier. At $\varepsilon =\pm 1$, only two eigenstates exist while all four eigenvalues coincide numerically. Beyond this value of $\varepsilon$, there are two pairs of identical complex-conjugate eigenvalues, and the associated basis is complete.   

For \ptc\ systems described by matrices of higher dimensions it is, in general, not possible to find the roots of the characteristic polynomial. Nevertheless, a discussion of the parameter space can still be given: to this end one needs to detect the number of real and complex eigenvalues for each set of parameter values; an algorithm capable of doing this will be presented in \cite{weigert05b}.  

\subsection*{\sf VI. Discussion and Outlook}\label{sec:Discussion-and-Outlook}

An algorithm has been presented which allows one to determine
whether a given PT-invariant matrix ${\sf M}$ is diagonalizable. To do so, it is not necessary to determine the roots of its characteristic 
polynomial. In terms of Linear Algebra, the algorithm decides whether the given matrix is {\em similar} \cite{lancaster+85} to a diagonal matrix or to a matrix containing at least one Jordan block of dimension two or more. Somewhat surprisingly, this question seems to have been addressed only recently from an algorithmic point of view. 

It seems worthwhile to point out that the test for multiple
roots of a polynomials can, in fact, be used without any change to
determine whether the eigenvalues of a given hermitean matrix are
degenerate or not. The present author is not aware that this 
observation has been made before.

When applied to a family of PT-symmetric matrices, the algorithm outputs polynomial conditions on the perturbation parameter. These conditions are satisfied
for sets of matrices all of which are not diagonalizable, and they
divide the full parameter space into regions of diagonalizable matrices
with qualitatively different spectra. When combined with an algorithm to identify the number of real and complex eigenvalues of {\sf M}, a complete picture of the system's properties in the entire parameter space can be established.    

Many PT-symmetric systems--including the first one studied from this perspective \cite{bender+98}--have been defined on Hilbert spaces with countably infinite dimension. Various concepts such
as eigenvalues and eigenstates, or the difference between algebraic and geometric multiplicities of degenerate eigenvalues continue to
exist in the more general case \cite{kato84}. In spite of a close similarity of hermitean operators in finite- and infinite-dimensional spaces, many concepts of the matrix case are not easily carried over to the more general situation. For any \emph{algorithm}, finiteness is a crucial feature: the number of steps required to identify a potential common factor of two polynomials
is always finite, no matter what their degree.  It will be interesting to see whether algorithmic tests for diagonalizability of operators acting on spaces with countably infinite dimension can be found.

\begin{thebibliography}{10}

\bibitem{bender04}
C. M. Bender: {\em Czech. J. Phys} {\bf 54}, 1027 (2004).

\bibitem{mostafazadeh04}
A. Mostafazadeh: {\em Czech. J. Phys} {\bf 54}, 1125 (2004).

\bibitem{weigert04}
St. Weigert:
\newblock {\em Czech. J. Phys} {\bf 54}, 1139 (2004).

\bibitem{kato84}
T.~Kato:
\newblock {\em Perturbation Theory for Linear Operators}
\newblock (Springer,New York 1984).

\bibitem{bender+98}
C.~M. Bender and S.~Boettcher.
\newblock {\em Phys. Rev. Lett.} {\bf 24}, 5243 (1998).

\bibitem{bender+02}
M.~V.~Berry C.~M.~Bender and A.~Mandilara:
\newblock {\em J. Phys. A} {\bf 35}, L467 (2002).

\bibitem{lancaster+85}
P. Lancaster and M. Tismenetsky:
\newblock {\em The Theory of Matrices}.
\newblock (Academic Press, Orlando $^2$1985).

\bibitem{abate97}
M. Abate:
\newblock {\em Amer. Math. Monthly} {\bf 104}, 824 (1997).

\bibitem{heiss04}
W. D. Heiss: {\em J. Phys. A} {\bf 37} 2455, (2004).

\bibitem{horn+99}
R.~A. Horn and Alexey~K. Lopatkin:
\newblock {\em Lin. Alg. Appl.} {\bf 299}, 153 (1999).

\bibitem{horn+85}
R.~A. Horn and C.~R. Johnson.
\newblock {\em Matrix Analysis}.
\newblock (Cambridge University Press, New York 1985).

\bibitem{weisstein+05}
E.~W.~Weisstein et~al.:
\newblock {\em Matrix minimal polynomial}; 
\newline URL = {\tt http://mathworld.wolfram.com/MatrixMinimalPolynomial.html}

\bibitem{weigert05b}
St. Weigert:
\newblock {\em An Algorithmic Test for the `Breaking' of PT-Invariance in
  Finite-Dimensional Systems} (in preparation)

\end{thebibliography}
\end{document}